# Proper Affine Vector Fields in Bianchi Type I Space-Times

Ghulam Shabbir

Faculty of Engineering Sciences

GIK Institute of Engineering Sciences and Technology

Topi Swabi, NWFP, Pakistan

Email: shabbir@giki.edu.pk

**Abstract:**

A study of Bianchi type I space-times according to its proper affine vector field is given by using holonomy and decomposability, the rank of the $6\times 6$ Rieman matrix and direct integration techinques. It is shown that the special class of the above space-times admits proper affine vector fields.

## 1. INTRODUCTION

Through out $M$ is representing the four dimensional, connected, hausdorff space-time manifold with Lorentz metric g of signature (-, +, +, +). The curvature tensor associated with g, through Levi-Civita connection $\Gamma$, is denoted in component form by $R^a{}_{bcd}$. The usual covariant, partial and Lie derivatives are denoted by a semicolon, a comma and the symbol L, respectively. Round and square brackets denote the usual symmetrization and skew-symmetrization, respectively. The space-time $M$ will be assumed nonflat in the sense that the Riemann tensor does not vanish over any non empty open subset of $M$.

A vector field $X$ on $M$ is called an affine vector field if it satisfies

$$X_{a;bc} = R_{abcd} X^d. \tag{1}$$

or equivalently

$$X_{a,bc} - \Gamma^f_{ac} X_{f,b} - \Gamma^f_{bc} X_{a,f} - \Gamma^e_{ab} X_{e,c} + \Gamma^e_{ab}\Gamma^f_{ec} X_f - (\Gamma^e_{ab})_{,c} X_e - \Gamma^f_{ab}\Gamma^e_{cf} X_e$$
$$+ \Gamma^e_{fb}\Gamma^f_{ca} X_e + \Gamma^e_{af}\Gamma^f_{bc} X_e = R_{abcd} X^d.$$

If one decomposes $X_{a;b}$ on $M$ into its symmetric and skew-symmetric parts

$$X_{a;b} = \frac{1}{2}h_{ab} + F_{ab} \qquad (h_{ab} = h_{ba},\ F_{ab} = -F_{ba}) \tag{2}$$

then equation (1) is equivalent to

$$(i)\ h_{ab;c} = 0 \quad (ii)\ F_{ab;c} = R_{abcd} X^d \quad (iii)\ F_{ab;c} X^c = 0. \qquad (3)$$

Such a vector field $X$ is called affine if the local diffeomorphisms $\varphi_t$ (for appropriate $t$) associated with $X$ map geodesics into geodesics. If $h_{ab} = 2c g_{ab}, c \in R$, then the vector field $X$ is called homothetic (and *Killing* if $c = 0$). The vector field $X$ is said to be proper affine if it is not homothetic vector field and also $X$ is said to be proper homothetic vector field if it is not Killing vector field on $M$ [1]. Define the subspace $S_p$ of the tangent space $T_p M$ to $M$ at $p$ as those $k \in T_p M$ satisfying

$$R_{abcd} k^d = 0. \qquad (4)$$

## 2. Affine Vector Fields

In this section we will briefly discuss when the space-times admit proper affine vector fields for further details see [2].

Suppose that $M$ is a simple connected space-time. Then the holonomy group of $M$ is a connected Lie subgroup of the idenity component of the Lorentz group and is thus characterized by its subalgebra in the Lorentz algebra. These have been labeled into fifteen types $R_1 - R_{15}$ [3,4]. It follows from [2] that the only such space-times which could admit proper affine vector fields are those which admit nowhere zero covariantly constant second order symmetric tensor field $h_{ab}$ and it is known that this forces the holonomy type to be either $R_2$, $R_3$, $R_4$, $R_6$, $R_7$, $R_8$, $R_{10}$, $R_{11}$ or $R_{13}$. Here, we will only discuss the space-times which has the holonomy type $R_2$, $R_4$, $R_7$, $R_{10}$ or $R_{13}$.

First consider the case when $M$ has type $R_{13}$. Then one can always set up local coordinates $(t, x^1, x^2, x^3)$ on an open set $U = U_1 \times U_2$, where $U_1$ is a one dimensional timelike submanifold of $U$ coordinatized by $t$ and $U_2$ is a three dimensional spacelike submanifold of $U$ coordinatized by $x^1, x^2, x^3$ and where the above product is a metric product and the metric on $U$ is given by [1]

$$ds^2 = -dt^2 + g_{\alpha\beta} dx^\alpha dx^\beta \qquad (\alpha, \beta = 1, 2, 3) \qquad (5)$$

where $g_{\alpha\beta}$ depends on $x^\gamma$, $(\gamma = 1, 2, 3)$. The above space-time is clearly 1+3 decomposable. The curvature rank of the above space-time is atmost three and there exists a unique nowhere zero vector field $t_a = t_{,a}$ satisfying $t_{a;b} = 0$ and also $t_a t^a = -1$. From the Ricci Identity $R^a{}_{bcd} t^d = 0$. It follows from [2] that affine vector fields in this case are

$$X = (c_1 t + c_2)\frac{\partial}{\partial t} + Y \tag{6}$$

where $c_1, c_2 \in R$ and $Y$ is a homothetic vector field in the induced geometry on each of the three dimensional submanifolds of constant $t$.

Now consider the situation when $M$ has type $R_{10}$. The situation is similar to that of previous $R_{13}$ case except that now we have local decomposition is $U = U_1 \times U_2$, where $U_1$ is a one dimensional spacelike submanifold of $U$ and $U_2$ is a three dimensional timelike submanifold of $U$. The space-time metric on $U$ is given by [1]

$$ds^2 = dx^2 + g_{\alpha\beta} dx^\alpha dx^\beta \qquad (\alpha, \beta = 0, 2, 3) \tag{7}$$

where $g_{\alpha\beta}$ depends on $x^\gamma$, $(\gamma = 0, 2, 3)$. The above space-time is clearly 1+3 decomposable. The curvature rank of the above space-time is atmost three and there exists a unique nowhere zero vector field $x_a = x_{,a}$ satisfying $x_{a;b} = 0$ and also $x_a x^a = 1$. From the Ricci Identity $R^a{}_{bcd} x^d = 0$. It follows from [2] that affine vector fields in this case are

$$X = (c_1 x + c_2)\frac{\partial}{\partial x} + Y \tag{8}$$

where $c_1, c_2 \in R$ and $Y$ is a homothetic vector field in the induced geometry on each of the three dimensional submanifolds of constant $x$.

Next suppose $M$ has type $R_7$. Then each $p \in M$ has a neighborhood $U$ which decomposes metrically as $U = U_1 \times U_2$, where $U_1$ is a two dimensional submanifold of $U$ with an induced metric of Lorentz signature and $U_2$ is a two dimensional submanifold of $U$ with positive definite induced metric. The space-time metric on $U$ is given by [1]

$$ds^2 = P_{AB} dx^A dx^B + Q_{\alpha\beta} dx^\alpha dx^\beta \tag{9}$$

where $P_{AB} = P_{AB}(x^C), \forall A,B,C = 0,1$ and $Q_{\alpha\beta} = Q_{\alpha\beta}(x^\gamma), \forall \alpha,\beta,\gamma = 2,3$ and the above space-time is clearly 2+2 decomposable. The space-time (8) admits two recurrent vector fields [5] $l$ and n i.e. $l_{a;b} = l_a p_b$ and $n_{a;b} = n_a p_b$ where $p_b$ is the recurrent 1-form. It also admits two covariantly constant second order symmetric tensors which are $2l_{(a}n_{b)}$ and $(x_a x_b + y_a y_b)$. The rank of the $6 \times 6$ Riemann matrix is two. It follows from [2] that if $X$ is an affine vector field on $M$ then $X$ decomposes as

$$X = X_1 + X_2 \tag{10}$$

where the vector fields $X_1$ and $X_2$ are tangent to the two dimensional timelike and spacelike submanifolds, respectively. It also follows from [2] that $X_1$ and $X_2$ are homothetic vector fields in their respective submanifolds with their induced geometry. Conversely, every pair of affine vector fields, one in the timelike submanifolds and one spacelike submanifolds give rise to a affine vector field in space-time.

Now suppose that $M$ has type $R_4$. Then each $p \in M$ has a neighborhood $U$ which decomposes metrically as $U = U_1 \times U_2 \times U_3$, where $U_1$ and $U_2$ are one dimensional submanifold of $U$ and $U_3$ is a two dimensional submanifold of $U$. The space-time metric on $U$ is given by [2]

$$ds^2 = -dt^2 + dx^2 + g_{AB} dx^A dx^B \qquad (A,B = 2,3) \tag{11}$$

where $g_{AB}$ depends only on $x^C$ $(C = 2,3)$. The above space-time is clearly 1+1+2 decomposable. The curvature rank of the above space-time is one and there exist two independent nowhere zero unit timelike and spacelike covariantly constant vector field $t_a = t_{,a}$ and $x_a = x_{,a}$ satisfying $t_{a;b} = 0$ and $x_{a;b} = 0$. From the Ricci identity $R^a{}_{bcd} t_a = 0$ and $R^a{}_{bcd} x_a = 0$. It follows from [2] that affine vector fields in this case are

$$X = (c_1 t + c_2 x + c_3)\frac{\partial}{\partial t} + (c_4 t + c_5 x + c_6)\frac{\partial}{\partial x} + Y \qquad (12)$$

where $c_1, c_2, c_3, c_4, c_5, c_6 \in R$ and $Y$ is a homothetic vector on each of two dimensional submanifolds of constant $t$ and $x$.

Now suppose that $M$ has type $R_2$. Here each $p \in M$ admits a neighborhood $U$ which decomposes metrically as $U = U_1 \times U_2 \times U_3$, where $U_1$ and $U_2$ are one dimensional submanifold of $U$ and $U_3$ is a two dimensional submanifold of $U$. The space-time metric on $U$ is given by [2]

$$ds^2 = dy^2 + dz^2 + g_{AB} dx^A dx^B \qquad (A, B = 0,1) \qquad (13)$$

where $g_{AB}$ depends only on $x^C$ $(C = 0,1)$. The above space-time is clearly 1+1+2 decomposable. The curvature rank of the above space-time is one and there exist two independent nowhere zero unit timelike and spacelike covariantly constant vector field $y_a = y_{,a}$ and $z_a = z_{,a}$ satisfying $y_{a;b} = 0$ and $z_{a;b} = 0$. From the Ricci identity $R^a{}_{bcd} y_a = 0$ and $R^a{}_{bcd} z_a = 0$. It follows from [2] that affine vector fields in this case are

$$X = (c_1 y + c_2 z + c_3)\frac{\partial}{\partial y} + (c_4 y + c_5 z + c_6)\frac{\partial}{\partial z} + Y \qquad (14)$$

where $c_1, c_2, c_3, c_4, c_5, c_6 \in R$ and $Y$ is a homothetic vector on each of two dimensional submanifolds of constant $y$ and $z$.

### 3. MAIN RESULTS

As mentioned in section 2, the space-times which admit proper affine vector fields having holonomy type $R_2$, $R_3$, $R_4$, $R_6$, $R_7$, $R_8$, $R_{10}$, $R_{11}$ or $R_{13}$. It also follows from [5] that the rank of the $6 \times 6$ Riemann matrix is atmost three. Here in this paper we will consider the rank of the $6 \times 6$ Riemann matrix to study affine vector fields in Bianchi type I space-times. Consider Bianchi type I space-time in the usual coordinate system $(t, x, y, z)$ with line elememt [6]

$$ds^2 = -dt^2 + f(t) dx^2 + k(t) dy^2 + h(t) dz^2 \qquad (15)$$

where $f, k$ and $h$ are some nowhere zero functions of t only. It follows from [7], the above space-time admits three independent Killing vector fields which are

$$\frac{\partial}{\partial t}, \frac{\partial}{\partial y}, \frac{\partial}{\partial z}.$$

The non-zero independent components of the Riemann tensor are

$$R_{1010} = -\frac{1}{4}\frac{2\ddot{f}f - \dot{f}^2}{f} = \alpha_1, \quad R_{2020} = -\frac{1}{4}\frac{2\ddot{k}k - \dot{k}^2}{k} = \alpha_2,$$

$$R_{3030} = -\frac{1}{4}\frac{2\ddot{h}h - \dot{h}^2}{h} = \alpha_3, \quad R_{2121} = \frac{\ddot{k}f}{4} = \alpha_4, \quad (16)$$

$$R_{3131} = \frac{\ddot{h}f}{4} = \alpha_5, \quad R_{2323} = \frac{\ddot{k}h}{4} = \alpha_6.$$

Writing the curvature tensor with components $R_{abcd}$ at p as a $6 \times 6$ symmetric matrix in a well known way [8]

$$R_{abcd} = diag(\alpha_1, \alpha_2, \alpha_3, \alpha_4, \alpha_5, \alpha_6) \quad (17)$$

where $\alpha_1, \alpha_2, \alpha_3, \alpha_4, \alpha_5$ and $\alpha_6$ are real functions of $t$. The 6-dimensional labeling is in the order 10, 20, 30, 21, 31, 23 with $x^0 = t$. We are only interested in those case when the rank of the $6 \times 6$ Riemann matrix is less than or equal to three (excluding the flat cases). We thus obtain the following cases:

(A1) Rank=3, $f \in R - \{0\}, k = k(t), h = h(t)$

(A2) Rank=3, $k \in R - \{0\}, f = f(t), h = h(t)$

(A3) Rank=3, $h \in R - \{0\}, k = k(t), f = f(t)$

(A4) Rank=3, $f = f(t), k = k(t), h = h(t), -2\ddot{k}k + \dot{k}^2 = 0, -2\ddot{h}h + \dot{h}^2 = 0,$
$-2\ddot{f}f + \dot{f}^2 = 0$

(B1) Rank=2, $f \in R - \{0\}, k = k(t), h = h(t), -2\ddot{k}k + \dot{k}^2 = 0$

(B2) Rank=2, $k \in R - \{0\}, f = f(t), h = h(t), -2\ddot{h}h + \dot{h}^2 = 0$

(B3) Rank=2, $h \in R - \{0\}, k = k(t), f = f(t), -2\ddot{f}f + \dot{f}^2 = 0$

(B4) Rank=2, $f \in R - \{0\}, k = k(t), h = h(t), -2\ddot{h}h + \dot{h}^2 = 0$

(B5) Rank=2, $k \in R - \{0\}, h = h(t), f = f(t), -2\ddot{f}f + \dot{f}^2 = 0$

(B6) Rank=2, $h \in R - \{0\}, k = k(t), f = f(t), -2\ddot{k}k + \dot{k}^2 = 0$

(C1) Rank=1, $f, h \in R - \{0\}, k = k(t)$

(C2) Rank=1, $k, h \in R - \{0\}, f = f(t)$

(C3) Rank=1, $f, k \in R - \{0\}, h = h(t)$

(D1) Rank=1, $f \in R - \{0\}, k = k(t), h = h(t), -2\ddot{k}k + \dot{k}^2 = 0, -2\ddot{h}h + \dot{h}^2 = 0$

(D2) Rank=1, $k \in R - \{0\}, f = f(t), h = h(t), -2\ddot{f}f + \dot{f}^2 = 0, -2\ddot{h}h + \dot{h}^2 = 0$

(D3) Rank=1, $h \in R - \{0\}, f = f(t), k = k(t), -2\ddot{f}f + \dot{f}^2 = 0, -2\ddot{k}k + \dot{k}^2 = 0.$

We consider each case in turn.

**Case A1**

In this case $f \in R - \{0\}, k = k(t), h = h(t)$ and the rank of the $6 \times 6$ Riemann matrix is 3 and there exists a unique (up to a multiple) nowhere zero spacelike vector field $x_a = x_{,a}$ such that $x_{a;b} = 0$ (and so, from the Ricci idenity $R^a{}_{bcd} x_a = 0$). The line element can, after a recaling of $x$, be written in the form

$$ds^2 = dx^2 + (-dt^2 + kdy^2 + hdz^2). \tag{18}$$

The above space-time is clearly 1+3 decomposable and its holonomy type is $R_{10}$. The affine vector fields in this case [2] are

$$X = (c_3 x + c_4)\frac{\partial}{\partial x} + X' \tag{19}$$

where $c_3, c_4 \in R$ and $X'$ is a homothetic vector field in the induced geometry on each of the three dimensional submanifolds of constant $x$. The completion of case A necessities finding an homothetic vector fields in the induced geometry of the submanifolds of constant x. The induced metric $g_{\alpha\beta}$ (where $\alpha, \beta = 0, 2, 3$) with nonzero components is given by

$$g_{11} = -1, g_{22} = k(t), g_{33} = h(t). \tag{20}$$

A vector field $X'$ is called homothetic vector field if it satisfies $L_{X'}g_{\alpha\beta} = 2cg_{\alpha\beta}$, where $c \in R$. One can expand by using (19) to get

$$X^0{}_{,0} = c \tag{21}$$

$$kX^2{}_{,0} - X^0{}_{,2} = 0 \tag{22}$$

$$hX^3{}_{,0} - X^0{}_{,3} = 0 \tag{23}$$

$$\dot{k}X^0 + 2kX^2{}_{,2} = 2c \tag{24}$$

$$kX^2{}_{,3} + hX^3{}_{,2} = 0 \tag{25}$$

$$\dot{h}X^1 + 2hX^3{}_{,3} = 2c. \tag{26}$$

Equations (21), (22) and (23) give

$$X^0 = ct + A^1(y,z), \quad X^2 = A^1{}_y(y,z)\int \frac{1}{k}dt + A^2(y,z)$$

$$X^3 = A^1{}_z(y,z)\int \frac{1}{h}dt + A^3(y,z)$$

where $A^1(y,z)$ $A^2(y,z)$ and $A^3(y,z)$ are functions of integration. If one proceed further after a strightforward calculation one can find that proper homothetic vector fields exist if and only if

$$k = (at+c)^2 \quad h = (bt+d)^2 \tag{27}$$

where $a,b,c,d \in R(a,b \neq 0)$. Substituting (27) in (16) one finds that the rank of $6\times 6$ Riemann matrix is reduces to one thus giving a contradiction (since we are assuming that that the rank of $6\times 6$ Riemann matrix is three). So the only homothetic vector fields in the induced gemetry are the Killing vector fields which are

$$X^0 = 0, \quad X^2 = c_1, \quad X^3 = c_2 \tag{28}$$

where $c_1, c_2 \in R$. The affine vector fields in this case are (from (19) and (28))

$$X^0 = 0, \quad X^1 = xc_3 + c_4, \quad X^2 = c_1, \quad X^3 = c_2. \tag{29}$$

Cases (A2) and (A3) are exactly same.

**Class A4**

In this case $f = f(t), k = k(t), h = h(t), -2\ddot{k}k + \dot{k}^2 = 0, -2\ddot{h}h + \dot{h}^2 = 0$ and $-2\ddot{f}f + \dot{f}^2 = 0$. Equations $-2\ddot{k}k + \dot{k}^2 = 0, -2\ddot{h}h + \dot{h}^2 = 0$ and $-2\ddot{f}f + \dot{f}^2 = 0$ $\Rightarrow k = (a_1 t + a_2)^2, h = (a_3 t + a_4)^2$ and $f = (a_5 t + a_6)^2$, respectivily, where $a_1, a_2, a_3, a_4, a_5, a_6 \in R (a_1, a_3, a_5 \neq 0)$. We first suppose that $a_1 \neq a_3$, $a_1 \neq a_5$, $a_3 \neq a_5$, $a_2 \neq a_4, a_2 \neq a_6$ and $a_4 \neq a_6$. The rank of the $6 \times 6$ Riemann matrix is 3 and there exists a unique (up to a multiple) solution $t_a = t_{,a}$ of equation (4) but $t_a$ is not covariantly constant. The line element is

$$ds^2 = -dt^2 + (a_5 t + a_6)^2 dx^2 + (a_1 t + a_2)^2 dy^2 + (a_3 t + a_4)^2 dz^2. \quad (30)$$

Affine vector fields in this case are

$$X^0 = 0, \ X^1 = c_3, \ X^2 = c_1, \ X^3 = c_2 \quad (31)$$

where $c_1, c_2, c_3 \in R$. Affine vector fields in this case are Killing vector fields.

Now suppose $a_1 = a_3 = a_5 = a(say) \in R (a \neq 0)$ and $a_2 = a_4 = a_6 = b \in R$. The line element is

$$ds^2 = -dt^2 + (at + b)^2 (dx^2 + dy^2 + dz^2). \quad (32)$$

It follows from [10] that affine vector fields in this case are

$$X^0 = c_7 t, \ X^1 = -c_2 y + c_3 z + c_4$$
$$X^2 = c_2 x - c_5 y + c_6, \ X^3 = -c_3 x + c_5 y + c_1 \quad (33)$$

where $c_1, c_2, c_3, c_4, c_5, c_6, c_7 \in R.$

**Class B1**

In this case $f \in R - \{0\}, k = k(t), h = h(t)$ and $-2\ddot{k}k + \dot{k}^2 = 0$. Equation $-2\ddot{k}k + \dot{k}^2 = 0 \Rightarrow k = (at + b)^2$ where $a, b \in R (a \neq 0)$. The rank of the $6 \times 6$ Riemann matrix is two and there exists a unique (up to a multiple) nowhere zero spacelike vector field $x_a = x_{,a}$ such that $x_{a;b} = 0$ (and so, from the Ricci idenity $R^a{}_{bcd} x_a = 0$). After a suitable recaling of $x$ the line element takes the form

$$ds^2 = dx^2 + (-dt^2 + (at + b)^2 dy^2 + h dz^2). \quad (34)$$

The above space-time is clearly 1+3 decomposable and its holonomy type is $R_{10}$. The affine vector fields in this case are of the form (19). The completion of case B1 necessities finding an homothetic vector fields in the induced geometry of the submanifolds of constant $x$. The induced metric $g_{\alpha\beta}$ (where $\alpha, \beta = 0, 2, 3$) with nonzero components is given by

$$g_{11} = -1, \; g_{22} = (at+b)^2, \; g_{33} = h(t). \tag{35}$$

A vector field $X'$ is called homothetic vector field if it satisfies $L_{X'} g_{\alpha\beta} = 2c g_{\alpha\beta}$, where $c \in R$. One can expand by using (35) to get

$$X^0{}_{,0} = c \tag{36}$$

$$(at+b)^2 X^2{}_{,0} - X^0{}_{,2} = 0 \tag{37}$$

$$h X^3{}_{,0} - X^0{}_{,3} = 0 \tag{38}$$

$$a(at+b) X^0 + (at+b)^2 X^2{}_{,2} = c \tag{39}$$

$$(at+b)^2 X^2{}_{,3} + h X^3{}_{,2} = 0 \tag{40}$$

$$\dot{h} X^1 + 2h X^3{}_{,3} = 2c. \tag{41}$$

Equations (36), (37) and (38) give

$$X^0 = ct + A^1(y,z), \; X^2 = -A^1{}_y(y,z) \frac{1}{a(at+b)} + A^2(y,z)$$

$$X^3 = A^1{}_z(y,z) \int \frac{1}{h} dt + A^3(y,z)$$

where $A^1(y,z)$ $A^2(y,z)$ and $A^3(y,z)$ are functions of integration. If one proceed further after a strightforward calculation one can find that proper homothetic vector fields exist if and only if

$$h = (et + f)^2 \tag{42}$$

where $e, f \in R(e \neq 0)$. Substituting (42) in (16) one finds that the rank of the $6 \times 6$ Riemann matrix is reduces to one thus giving a contradiction (since we are assuming that that the rank of the $6 \times 6$ Riemann matrix is two). So the only homothetic vector fields in the induced gemetry are the Killing vector fields which are given in (28). The affine vector fields in this case are given in (29). Cases (B2), (B3), (B4), (B5) and (B6) are exactly same.

**Case C1**

In this case $f, h \in R - \{0\}, k = k(t)$ and the rank of the $6 \times 6$ Riemann matrix is one. There exist two independent nowhere zero spacelike vector field $x_a = x_{,a}$ and $z_a = z_{,a}$ satisfying $x_{a;b} = 0$ and $z_{a;b} = 0$. From the Ricci identity $R^a{}_{bcd} x_a = R^a{}_{bcd} z_a = 0$. The line element can, after a rescaling of x and z, be written as

$$ds^2 = dx^2 + dz^2 + (-dt^2 + k\, dy^2) \tag{43}$$

The above space-time is 1+1+2 decomposable and its holonomy type is $R_2$. Affine vector fields in this case are [2]

$$X = (c_1 x + c_2 z + c_3)\frac{\partial}{\partial x} + (c_4 x + c_5 z + c_6)\frac{\partial}{\partial z} + X' \tag{44}$$

where $c_1, c_2, c_3, c_4, c_5, c_6 \in R$ and $X'$ is a homothetic vector fields on each of two dimensional submanifolds of constant x and z. The next step is to find the homothetic vector fields in the induced geometry of the submanifolds of constant x and z. The induced metric $g_{AB}$ (where $A, B = 0, 2$) with non zero components is given by

$$g_{00} = -1, \quad g_{22} = k. \tag{45}$$

A vector field $X'$ is called homothetic vector field if it satisfies $L_{X'} g_{AB} = 2c g_{AB}$, where $c \in R$. One can expand by using (45) to get

$$X^0{}_{,0} = c \tag{46}$$

$$k X^2{}_{,0} - X^0{}_{,2} = 0 \tag{47}$$

$$\dot{k} X^0 + 2k X^2{}_{,2} = 2c k. \tag{48}$$

Equation (46) gives $X^0 = ct + A^1(y)$, where $A^1(y)$ is a function of integration. Using $X^0$ in equation (47) we get $X^2 = A^1_y(y) \int \frac{1}{k} dt + A^2(y)$, where $A^2(y)$ is a function of integration. If one proceeds further, after a straightforward calculation one finds that proper homothetic vector field exist if and only if $k = at^2$, where $a \in R - \{0\}$. Substituting the value of $k$ into (16), one finds that the rank of $6 \times 6$ Riemann matrix reduces to zero thus giving a contradiction (since we are assuming that the rank of $6 \times 6$ Riemann matrix is one). So homothetic vector fields in the induced geometry of constant $y$ and $z$ are Killing vector fields. If one proceeds further one finds there exist two possibilities:

$$\text{(a)}\ k\left(\frac{\dot{k}}{2k}\right)^{\cdot} = n \qquad \text{(b)}\ k\left(\frac{\dot{k}}{2k}\right)^{\cdot} \neq n$$

where $n \in R$.

**Case C1a**

In this case further three possibilities exist

(i) $n > 0$, (ii) $n < 0$, (iii) $n = 0$.

We will consider each case in turn.

**(i)** Affine vector fields in this case are

$$\begin{aligned}
X^0 &= c_7 \sin \sqrt{n}\ y + c_8 \cos \sqrt{n}\ y, \\
X^1 &= c_1 y + c_2 z + c_3, \\
X^2 &= \sqrt{n}\ (c_7 \cos \sqrt{n}\ y - c_8 \sin \sqrt{n}\ y) \int \frac{1}{k} dt + c_9, \\
X^3 &= c_4 y + c_5 z + c_6.
\end{aligned} \tag{49}$$

provided that $k\left(\dfrac{\dot{k}}{2k}\right)^{\cdot} = n$. Where $c_7, c_8, c_9 \in R$.

**(ii)** In this case $n < 0$. Put $n = -N$, where $N \in R(N > 0)$. Affine vector fields in this case are

$$\begin{aligned}
X^0 &= c_7 \sinh \sqrt{N}\ y + c_8 \cosh \sqrt{N}\ y, \\
X^1 &= c_1 y + c_2 z + c_3, \\
X^2 &= \sqrt{N}\ (c_7 \cosh \sqrt{N}\ y - c_8 \sinh \sqrt{N}\ y) \int \frac{1}{k} dt + c_9, \\
X^3 &= c_4 y + c_5 z + c_6.
\end{aligned} \tag{50}$$

provided that $k\left(\dfrac{\dot{k}}{2k}\right)^{\cdot} = -N$. Where $c_7, c_8, c_9 \in R$.

**(iii)** In this case $n = 0 \Rightarrow k = e^{at+b}$, where $a, b \in R(a \neq 0)$. Affine vector fields in this case are

$$X^0 = c_7 y + c_8, \qquad X^1 = c_1 y + c_2 z + c_3,$$
$$X^2 = -\frac{c_7}{a} e^{-(at+b)} + c_9, \qquad X^3 = c_4 y + c_5 z + c_6. \qquad (51)$$

where $c_7, c_8, c_9 \in R$.

**Case C1b**

Affine vector fields in this case are

$$X^0 = c_7, \qquad X^1 = c_1 y + c_2 z + c_3, \qquad X^2 = 0,$$
$$X^3 = c_4 y + c_5 z + c_6. \qquad (52)$$

where $c_i \in R$. This completes case C1. Cases C2 and C3 are exactly same.

**Case D1**

In this case $f \in \Re - \{0\}, k = k(t), h = h(t)$, $-2\ddot{k}k + \dot{k}^2 = 0$ and $-2\ddot{h}h + \dot{h}^2 = 0$. Equations $-2\ddot{k}k + \dot{k}^2 = 0$ and $-2\ddot{h}h + \dot{h}^2 = 0 \Rightarrow k = (at+b)^2$ and $h = (ct+b)^2$, respectively where $a,b,c,d \in R(a \neq c, a, c \neq 0)$. The rank of the $6 \times 6$ Riemann matrix is 1 and there exist two independent nowhere zero solutions $t_a = t_{,a}$ and $x_a = x_{,a}$ of equation (4) with $t$ timelike and $x$ spacelike vector fields, respectivily and $x_{a;b} = 0$. After a rescaling of $x$ the line element is

$$ds^2 = -dt^2 + dx^2 + (at+b)^2 dy^2 + (ct+d)^2 dz^2. \qquad (53)$$

The space-time is clearly 1+3 decomposable and the rank of the $6 \times 6$ Riemann matrix is 1. Substituting the above information into affine equations and after a strightforward calculation one find affine vector fields in this case are

$$X^0 = 0, \quad X^1 = c_3 t + c_4 x + c_5, \quad X^2 = c_1, \quad X^3 = c_2 \qquad (54)$$

where $c_1, c_2, c_3, c_4, c_5 \in R$.

Now suppose $a = c, b = d \in R(a, c \neq 0)$. The line element is

$$ds^2 = -dt^2 + dx^2 + (at+b)^2 (dy^2 + dz^2). \qquad (55)$$

We know from [8,9] that affine vector fields in this case are

$$X^0 = c_4(at+b), \quad X^1 = c_5 x + c_6, \quad X^2 = -c_3 z + c_1, \quad X^3 = c_3 y + c_2 \qquad (56)$$

where $c_1, c_2, c_3, c_4, c_5, c_6 \in R$. Cases (D2) and (D3) are exactly the same.

# SUMMARY

In this paper a study of Bianchi type I space-times according to their proper affine vector fields is given. An approach is developed to study proper affine vector fields in the above space-times by using the rank of the $6\times6$ Riemann matrix and holonomy. From the above study we obtain the following results:

(i)     The case when the rank of the $6\times6$ Riemann matrix is three and there exists a nowhere zero independent spacelike vector field which is the solution of equation (4) and is not covariantly constant. This is the space-time (30) and it admits affine vector fields which are Killing vector fields (for details see Case A4).

(ii)    The case when the rank of the $6\times6$ Riemann matrix is one and there exist two nowhere zero independent solution of equation (4) but only one independent covariantly constant vector field. . This is the space-time (55) and it admits proper affine vector fields (see Case D1).

(iii)   The case when the rank of the $6\times6$ Riemann matrix is two or three and there exists a nowhere zero independent spacelike vector field which is the solution of equation (4) and also covariantly constant. This is the space-time (18) and (34) and it admits proper affine vector fields (see Cases A1 and B1).

(iv)    In the case when the rank of the $6\times6$ Riemann matrix one there exist two nowhere zero independent spacelike and timelike vector fields which are solutions of equation (4) and are covariantly constant. This is the space-time (43) and it admits proper affine vector fields (see Case C1).

(v)     The case when the rank of the $6\times6$ Riemann matrix is three and there exists a nowhere zero independent spacelike vector field which is the solution of equation (4) and is not covariantly constant. This is the space-time (32) and it admits proper affine vector fields (see equation (33)).